\documentclass[12pt]{article}
\usepackage{graphicx}

\input{epsf}

\newcommand{\nc}{\newcommand}
\nc{\beq}{\begin{equation}}
\nc{\eeq}{\end{equation}}
\nc{\beqa}{\begin{eqnarray}}
\nc{\eeqa}{\end{eqnarray}}

\def\gsim{\mathrel{\rlap{\lower4pt\hbox{\hskip1pt$\sim$}}
    \raise1pt\hbox{$>$}}}       

\begin{document}



\title{\large{\bf Noncommutative General Relativity}}

\author{Xavier~Calmet$^1$\thanks{calmet@physics.unc.edu} \ and Archil~Kobakhidze$^2$\thanks{kobakhid@physics.unc.edu} \\
$^1$Universit\'e Libre de Bruxelles \\ 
Boulevard du Triomphe  (Campus plaine)\\
B-1050 Brussels, Belgium\\
$^2$Department of Physics and Astronomy \\
University of North Carolina at Chapel Hill \\
Chapel Hill, NC 27599, USA}

\date{June, 2005}

\maketitle

\begin{abstract}
We define a theory of noncommutative general relativity for canonical noncommutative spaces. We find a subclass of general coordinate transformations acting on canonical noncommutative spacetimes to be volume-preserving transformations. Local Lorentz invariance is treated as a gauge theory with the spin connection field taken in the so(3,1) enveloping algebra. The resulting theory appears to be a noncommutative extension of the unimodular theory of gravitation. We compute the leading order noncommutative correction to the action and derive the noncommutative correction to the equations of motion of the weak gravitation field.
\end{abstract}


\newpage

\section{Introduction}
General Relativity \cite{Einstein:1916vd} is a very successful theory when it comes to describe macroscopic effects of gravitation. However, it is widely believed that an unification of Quantum Mechanics and General Relativity requires a short distance modification of spacetime. It can be shown that classical General Relativity considered together with Quantum Mechanics implies the existence of a fundamental length \cite{Calmet:2004mp}. A class of models that incorporate the notion of a fundamental length in gauge theories are gauge theories formulated on noncommutative spaces. 

It is a challenge to formulate General Relativity on noncommutative spaces and there are thus different approaches in the literature. In \cite{Chamseddine:2000si} for example  a  deformation of Einstein's gravity was studied using a  construction  based on gauging the noncommutative SO(4,1) de Sitter group and the Seiberg-Witten map with subsequent contraction to ISO(3,1). Most recently another construction of a noncommutative gravitational theory \cite{Aschieri:2005yw} was proposed based on  a twisted Poincar\'e  algebra \cite{Wess:2004da,Chaichian:2004za}. These approaches although  mathematically consistent, are not minimal formulations of Einstein's General Relativity on noncommutative spaces.  The main problem in formulating a theory of gravity  on noncommutative manifolds is that  it is difficult to implement symmetries such as general coordinate covariance and local Lorentz invariance and to define derivatives which are torsion-free and satisfy the metricity condition. 

Similar obstacles appear in constructing models of particle physics on flat spacetime with canonical noncommutativity defined by the algebra $[ \hat x^a, \hat x^b ]=i \theta^{a b}$ ($\theta^{ab}$ is constant and antisymmetric). Indeed, it turns out to be rather difficult to implement most symmetries particle physicists are so familiar with. In particular, Lorentz invariance is explicitly violated by the noncommutative algebra. However, it has been shown in \cite{Calmet:2004ii} that there is another exact symmetry, the noncommutative Lorentz invariance, based on the usual Lorentz algebra so(3,1) which is undeformed. Another example is the implementation of noncommutative local gauge theories. A formulation of noncommutative gauge theories within the enveloping algebra approach has been proposed in \cite{Jurco:2000ja}.    
The fields taken in the enveloping algebra are expanded in term of a series in $\theta$. Each of the terms of this series is however a function of the classical variables \cite{Madore:2000en,Seiberg:1999vs}. The number of degrees of freedom is thus finite and the same as in the corresponding commutative gauge theory. 

In this work, based partially on the above achievements in implementing symmetries on flat noncommutative spacetimes, we would like to propose a theory of General Relativity on curved spacetimes with canonical noncommutativity. We shall use the tetrad approach to General Relativity. This formalism applied to noncommutative General Relativity allows to follow closely the usual construction of noncommutative gauge theories. This requires to implement two gauge symmetries: local Lorentz transformations which can be seen as a local gauge theory based on the algebra so(3,1) for the spin connection field and general coordinate transformations which are inhomogenous  translations with the tetrad as a gauge field. 

The gauging of noncommutative so(3,1) algebra is only possible if the corresponding gauge field, the spin connection, is assumed to be in the enveloping algebra. Hence, to implement local Lorentz invariance we follow  the approach developed in \cite{Jurco:2000ja}. The invariance under the general coordinate transformations, however, is explicitly violated by  the canonical noncommutative algebra. Nevertheless, we find a restricted class of coordinate transformations which preserve the canonical structure. It turns out that this transformations correspond to volume-preserving diffeomorphisms. Thus, the basic new ideas for constructing a theory of noncommutative General Relativity are to formulate local Lorentz invariance by gauging so(3,1) within the enveloping algebra approach and to introduce volume-preserving coordinate transformations in place of general coordinate transformations, which is indeed  an exact symmetry of the canonical noncommutative spacetime.

\section{Noncommutative General Relativity}
Let us start from a noncommutative  spacetime and assume that the coordinates fulfill canonical commutation relations:
\begin{eqnarray} 
[ \hat x^\mu, \hat x^\nu ]=i \theta^{\mu\nu}.
\label{a}
\end{eqnarray}
Obviously, the commutator (\ref{a}) explicitly violates general coordinate covariance since $\theta^{\mu\nu}$ is constant in all reference frames. However, we can identify a subclass of general coordinate transformations,
\begin{equation}
\hat x^{\mu \prime}=\hat x^{\mu}+\hat \xi^{\mu}(\hat x),
\label{c}
\end{equation}
which are compatible with the algebra given by (\ref{a}). The hat on the function $\hat \xi(\hat x)$ indicates that it is in the enveloping algebra.  Under the change of coordinates (\ref{c}) the commutator (\ref{a}) transforms as:
\begin{eqnarray}
[\hat x^{\mu \prime}, \hat x^{\nu \prime}]&=&
\hat x^{\mu \prime}\hat x^{\nu \prime}- \hat x^{\nu \prime} \hat x^{\mu \prime} 
=i \theta^{\mu\nu} + [\hat x^\mu, \hat \xi^\nu] + [\hat \xi^\mu, \hat x^\nu] + {\cal O}(\hat \xi^2) 
\label{d}
\end{eqnarray} 
Requiring that $\theta$ remains constant yields the following partial differential equations:
\begin{eqnarray}
\theta^{\mu \alpha} \hat \partial_{\alpha} \hat  \xi^\nu(\hat x) = \theta^{\nu \beta} \hat \partial_{\beta} \hat \xi^\mu(\hat x).
\end{eqnarray}
A nontrivial solution to this condition can be easily found:
\begin{equation}
\hat \xi^{\mu}(\hat x)=\theta^{\mu\nu}\hat \partial_{\nu} \hat f(\hat x),
\label{e}
\end{equation}
where $\hat f(\hat x)$ is an arbitrary field. This noncommutative general coordinate transformation corresponds to the following classical transformation: $\hat \xi^{\mu}(x)=\theta^{\mu\nu} \partial_{\nu} \hat f(x)$. The Jacobian of this restricted coordinate transformations is equal to 1, meaning that the volume element is invariant: $d^{4}x^{\prime}=d^{4}x$. The version of General Relativity based on volume-preserving diffeomorphism is known as the unimodular theory of gravitation \cite{UNI}. Thus we came to the conclusion that symmetries of canonical noncommutative spacetime naturally lead to the noncommutative version of unimodular gravity.

Now we need to implement two gauge symmetries mentioned above. A noncommutative gauge transformation $\hat \Lambda(\hat x)$ valued in the iso(3,1) Lie algebra can be decomposed using the generators of the inhomogeneous  translations $p_\mu= -i \partial_\mu$, which are anti-Hermitian, and the generators of the Local Lorentz algebra so(3,1) $\Sigma_{ab}$, which are Hermitian. One finds
\begin{eqnarray} \label{NCGaugeT}
\hat \Lambda(\hat x) =\hat \xi (\hat x) + \hat \Lambda(\hat x)= \hat \xi^\mu (\hat x) p_\mu + \frac{1}{2} \hat \lambda^{ab}(\hat x) \Sigma_{ab},
\end{eqnarray}
where $\hat \xi^{\mu}$ is subject to the constraint (\ref{e}). Note that $p_{\mu}$ acts on the coordinates and functions, including $\hat \lambda^{ab}$, and $a,b,...$ run over the tangent space indices. As in  \cite{Calmet:2004ii}, the algebra of generators is undeformed. It is easy to verify that the commutator of two noncommutative gauge transformations $[\hat \Lambda_1(\hat x),\hat \Lambda_2(\hat x)]$ is in general not a noncommutative gauge transformation if the transformations are Lie algebra valued. As in the Yang-Mills case, the solution is to assume that the noncommutative gauge transformations are in the enveloping algebra.
Let us introduce a noncommutative vector potential which corresponds to the noncommutative gauge transformation  (\ref{NCGaugeT})
\begin{eqnarray} \label{NCGaugeP}
\hat A_a (\hat x) = (\hat D_a)= i \hat E^\mu_a(\hat x) p_\mu + \frac{i}{2} \hat \omega(\hat x)_a^{\ bc} \Sigma_{bc}
\end{eqnarray}
where $ \hat E^\mu_a(\hat x)$ are the components of the noncommutative tetrad $ \hat E_a(\hat x)$ , i.e. the gauge fields  corresponding to general coordinate transformations and $\hat \omega(\hat x)_a^{\ bc}$ are the spin connections fields associated with local Lorentz invariance. Note that $\hat A_a (\hat x)$ plays a dual role. It can be viewed as a covariant  derivative as well. It is also worth noticing that $\hat E_a=\hat E_a^\mu \hat \partial_\mu=\hat \partial_a$, which implies that the noncommutative tetrad is mapped trivially on the commutative one: $\hat E_a= e_a$ to all orders in $\theta$.

Let us now assume that the gauge transformations and the spin connection field are in the enveloping algebra: 
\begin{eqnarray}
\hat \Lambda = \Lambda(x) + \Lambda^{(1)}(x, \omega_a)+ {\cal O} (\theta^2),
\end{eqnarray}
and
\begin{eqnarray}
\hat \omega_a = \omega_a(x) + \omega^{(1)}_a(x, \omega_a)+ {\cal O} (\theta^2),
\end{eqnarray}
respectively, with $\Lambda(x)= \xi^\mu(x)p_\mu +\frac{1}{2} \lambda^{ab}(x)\Sigma_{ab}$ and $\omega_a(x)=\frac{1}{2}
\omega_a^{\ bc}\Sigma_{bc}$. We require that the commutator of two noncommutative  gauge transformations with $\hat{\Lambda}_1$ and $\hat{\Lambda}_2$ be a gauge transformation $\hat{\Lambda}_{\widehat{\Lambda_1 \times \Lambda_2}}$:
\begin{eqnarray}
\left ( \hat \delta_{\hat \Lambda_1} \hat \delta_{\hat \Lambda_2} -
 \hat \delta_{\hat \Lambda_2} \hat \delta_{\hat \Lambda_1} 
 \right) \star \hat \phi(x) &=&
\left ( i \hat \delta_{\hat \Lambda_1} \hat \Lambda_2[\omega_a]  
- i \hat \delta_{\hat \Lambda_2} \hat \Lambda_1[\omega_a] + [\hat \Lambda_1[\omega_a] \stackrel{\star}{,} 
\hat \Lambda_2[\omega_a] \right ) \star \hat \phi(x)  \\ \nonumber
&=& \hat \Lambda_{\widehat{\Lambda_1 \times \Lambda_2}} \star \hat \phi( x).
\end{eqnarray}
One finds as usual
\begin{eqnarray}
[\Lambda_1,\Lambda_2]= i \Lambda_{\Lambda_1 \times \Lambda_2}
\end{eqnarray}
in the zeroth order in $\theta$ and 
\begin{eqnarray}
i \delta_{\Lambda_1} \Lambda^{(1)}_2 - i \delta_{\Lambda_2} \Lambda^{(1)}_1 
+ i \theta^{ab} \{\partial_a \Lambda_1, \partial_b \Lambda_2 \} + [\Lambda_1, \Lambda^{(1)}_2] - [\Lambda_2, 
\Lambda^{(1)}_1] = \Lambda^{(1)}_{\Lambda_1 \times \Lambda_2}
\end{eqnarray}
in the leading order in $\theta$.
A solution to this consistency equation is 
\begin{eqnarray}
\Lambda^{(1)}_1=\frac{1}{4} \theta^{ab} \{\partial_a \Lambda_1,\omega_b\}
\label{k}
\end{eqnarray}
and analogously for $\Lambda^{(1)}_2$ and $\Lambda^{(1)}_{\Lambda_1 \times \Lambda_2}$ and where we have used: $\theta^{ab}= \theta^{\mu\nu} e_\mu^a e_\nu^b$ and $\partial_a = e^\mu_a \partial_\mu$. Note that the consistency condition is derived in the leading order in $\theta$ and $\xi(x)$. However since $\xi(x)$ is itself proportional to $\theta$, the relevant part of the volume-preserving diffeomorphism transformation is trivial. In other words,  the terms proportional to  $\xi(x)$ can be dropped in equation (\ref{k}) and it actually determines $\lambda^{(1)}$ which is the first non-trivial term in the Seiberg-Witten map for the so(3,1) gauge transformation.

 The consistency condition for the spin connection is given by
 \begin{eqnarray}
\delta_\Lambda \omega^{(1)}_a&=&\partial_a \Lambda^{(1)} -\frac{1}{2} \theta^{bc} 
\left ( \partial_b \lambda \partial_c \omega_a - \partial_b \omega_a \partial_c \Lambda \right)
+[\Lambda, \omega^{(1)}_a] + i [\Lambda^{(1)}, \omega_a].
\end{eqnarray}
A solution is 
\begin{eqnarray}
\omega^{(1)}_a= -\frac{1}{4} \theta^{bc}\{\omega_b, \partial_c \omega_a + F_{ca} \}
\end{eqnarray}
One thus has $\hat E^\mu_a=e^\mu_a$ to all orders in $\theta$ and $\hat \omega_a = \omega_a + \omega^{(1)}_a+ {\cal O} (\theta^2)$. 

The Seiberg-Witten map for the field strength is given by $\hat F_{ab}= F_{ab} + F^{(1)}_{ab}+ {\cal O} (\theta^2)$, where
\begin{eqnarray}
F^{(1)}_{ab} =\frac{1}{2} \theta^{cd} \{F_{ac},F_{bd} \} -\frac{1}{4} \theta^{cd} \{\omega_c, (\partial_d + D_d) F_{ab} \}, 
\label{aa}
\end{eqnarray}
where $D_a=A_a= i e^\mu_a p_\mu +\frac{i}{2} \omega_a^{\ bc} \Sigma_{bc}$. is the commutative covariant derivative.
 
The commutative field strength $F_{ab}$ contains the Riemann tensor $R_{ab}^{\ \ cd}$ as well as  a torsion $T_{ab}^{\ \ c}$:
\begin{eqnarray}
F_{ab}= i [D_a, D_b]= \frac{1}{2} R_{ab}^{\ \ cd} \Sigma_{cd} + T_{ab}^{\ \ c} D_c
\label{bb}
\end{eqnarray}
with $R_{ab}=\frac{1}{2} R_{ab}^{\ \ cd} \Sigma_{cd}$ and $T_{ab}^{\ \ c}= (D_a e^\nu_b -D_b e^\nu_a) e^c_\nu$. The commutative covariant derivative $D_a$ is torsion free ($T_{ab}^{\ \ c}=0$) and compatible with a metric: $e^a_\mu D_a e^b_\nu=0$. We now have all the required tools to consider actions that are invariant under general coordinate transformations.

\section{Action for noncommutative General Relativity}
 The Seiberg-Witten map for the Riemann tensor $R_{ab}$ which is the field strength tensor corresponding to a local noncommutative Lorentz transformation can be read from equations (\ref{aa}) and (\ref{bb}) where the classical torsion is being set to zero:
\begin{eqnarray}
\hat R_{ab}=R_{ab} + R^{(1)}_{ab} + { \cal  O}(\theta^2),
\end{eqnarray}
with
\begin{eqnarray}
R^{(1)}_{ab} =\frac{1}{2} \theta^{cd} \{R_{ac},R_{bd} \} -\frac{1}{4} \theta^{cd} \{\omega_c, (\partial_d + D_d) R_{ab} \}.
\end{eqnarray}
The noncommutative Riemann tensor is then given by
\begin{eqnarray}
\hat R_{ab}(\hat x) =\frac{1}{2}\hat R_{ab}^{\ \ cd}(\hat x) \Sigma_{cd},
\end{eqnarray}
from which we can determine the corresponding noncommutative Ricci tensor, $\hat R_{ab}^{\ \ bd}$, 
and a Ricci scalar $\hat R=\hat R_{ab}^{\ \ ab}$ in terms of the classical fields using the above Seiberg-Witten map. 
 
The noncommutative action is then given by
\begin{eqnarray} \label{NCaction}
S&=&\int  d^4 x \frac{1}{2 \kappa^2} \hat R(\hat x) 
=
\int  d^4 x \frac{1}{2 \kappa^2} \left (R(x)+ R^{(1)}(x)\right)+{\cal O}( \theta^2).
\end{eqnarray}
In the second line we have made use of the Weyl quantization procedure which allows to replace the noncommutative variables by commuting ones by expanding the noncommutative fields using the Seiberg-Witten maps. The only correction to leading order in $\theta$ comes from the Seiberg-Witten map for the so(3,1) gauge field.
It is easy to verify that this action is Hermitian and invariant under unimodular coordinate transformations and local Lorentz transformations. This noncommutative general relativity theory is, by construction, torsion free. In the leading order in the expansion in $\theta$ we can use the classical relations:
\begin{eqnarray}
\omega_\mu^{ab}(x) = \frac{1}{2} e^c_\mu(x)  \left ( \Omega^{ab}_{\ \ c}(x) -\Omega^{b \ a}_{\ c}(x)   
 -\Omega^{\ ab}_{c}(x)  
\right )
\end{eqnarray}
with 
\begin{eqnarray}
\Omega^{ab}_{\ \ c}(x)  = e_\mu^ a(x)  e_\nu^b(x)  \left (\partial^\mu e_c^\nu(x)  -\partial^\nu e_c^\mu(x)  \right).
\end{eqnarray}
The equation (\ref{NCaction}) represents an action for the noncommutative version of the unimodular theory of gravitation. The unimodular theory is known \cite{UNI} to be classically equivalent to Einstein's General Relativity with a cosmological constant. Indeed, we can rewrite the action (\ref{NCaction}) in the form of an Einstein-Hilbert action by introducing a Lagrange multiplier $\Lambda$ which appears to be an arbitrary integration constant:
\begin{equation} \label{exaction}
S= \int  d^4x \left ( \frac{1}{2 \kappa^2}   \mbox{det} (e^a_\mu(x))\left(R(x)+ R^{(1)}(x)\right)+\Lambda \left(\mbox{det}(e^a_\mu(x))-1\right)
+ {\cal O}( \theta^2) \right).
\end{equation}
In deriving (\ref{exaction}) we have used:
\begin{eqnarray}
\mbox{det}_{\star}(e^a_\mu(x))\stackrel{\rm{def}}{=}\frac{1}{4!}\epsilon^{\mu \nu \rho \sigma}
\epsilon_{abcd}e^a_\mu(x)\star e^b_\nu(x)\star e^c_\rho(x)\star e^d_\sigma(x)=\mbox{det}(e^a_\mu(x))+ {\cal O}( \theta^2),
\end{eqnarray} 
where $\star$ is the star product.
 
Let us now consider the weak field approximation of the noncommutative action (\ref{exaction}). Although the theory defined by the action (\ref{exaction}) does not admit flat spacetime as a background solution, we can still locally (for regions with volume $V<<1/\Lambda$) expand the tetrad around  flat spacetime: 
\begin{eqnarray}
e^\mu_a(x)=\eta^\mu_a-\frac{1}{2}h^\mu_a(x). 
\end{eqnarray}
Here $h^\mu_a(x)$ is a weak gravitational field subject to the traceless condition ($\mbox{det}(e^a_\mu(x))=1$ follows from (\ref{exaction})).
The noncommutative correction to Einstein's action in the weak field limit reads
\begin{eqnarray}
\frac{1}{8} \theta^{ln}  \tilde R^{\ \ ab}_{kl}  \tilde R_{mn}^{\ \ cd} d_{abcd}^{km} -\frac{1}{16} \theta^{ln} \tilde R_{ln}^{\ \ ab} \tilde R_{km}^{\ \ cd} d_{abcd}^{km}
\end{eqnarray}
where $d_{abcd}^{km}$ are the structure constants defined by $d_{abcd}^{km}\Sigma_{km}=2 \{\Sigma_{ab}, \Sigma_{cd} \}$. Notice that 
\begin{eqnarray}
\tilde R_{km}^{\ \ ab}=-\frac{1}{2} \partial_k (\partial^a h^b_m +\frac{1}{2} \partial^b h^a_m) -  
\partial_m (\partial^a h^b_k - \partial^b h^a_k) 
\end{eqnarray}
is the leading order of the weak field approximation of $R_{ab}^{\ \ cd}$. This modification of the linearized noncommutative action implies a noncommutative correction of the equations of motion for the weak gravitational field:
\begin{eqnarray}
R^a_{\ b} -\frac{1}{2} \eta^a_{\ b}R= \frac{1}{8} \theta^{an} \left (\partial^r \partial_k \tilde R^{\ \ cd}_{mn}  + \partial^r \partial_n \tilde R^{\ \ cd}_{km} \right) d_{rbcd}^{km}
-\frac{1}{8} \theta^{ln} \partial^r \partial_l \tilde R_{mn}^{\ \ cd} d_{rbcd}^{\ \ am}
\end{eqnarray}
where $R^a_{\ b}$ is the usual Ricci tensor and $R$ is the corresponding Ricci scalar in the linearized approximation (we have omitted the contribution coming from the cosmological constant). This modification might have some interesting physical implications that will be studied elsewhere.

 Finally we briefly discuss the relation between the tetrad formalism considered here and a second-order formalism which involves the metric tensor. The noncommutative metric tensor defined naively as $\hat g_{\mu\nu}(\hat x)=E_\mu^a(\hat x) 
E_\nu^b(\hat x)\eta_{ab}$ is neither real nor symmetric. This raises a more generic question of the geometrical interpretation of the noncommutative deformation of Einstein's General Relativity considered above. The simple prescription to define the metric in our case is to solve the deformed equations of motion for the classical tetrads at each given order of $\theta$ expansion and then to determine the metric in the standard way.

\section{Conclusions}
We  have constructed a theory of noncommutative General Relativity on a canonical noncommutative spacetime. The general coordinate transformations is shown to be restricted to the volume-preserving transformations. Thus the General Relativity on canonical noncommutative spacetimes is the noncommutative version of the unimodular theory of gravitation. The local Lorentz invariance is described as a noncommutative gauge theory by taking the spin-connection field in the enveloping algebra.

The action for noncommutative General Relativity was constructed and the expansion in first order of the noncommutative parameter $\theta$ has been calculated. We derived the noncommutative correction to the equations of motion of the weak gravitational field. An interesting question is whether the effects  of spacetime noncommutativity  coming from the noncommutative modifications of gravity  are stronger than the ones coming from the modifications of the interactions of the standard model \cite{Calmet:2004dn}.

It will also be interesting to consider classical solutions of the noncommutative action defined in this work. The minimal length introduced in our version of General Relativity could have interesting consequences for the horizon and the singularity of black holes. It will also be worth  studying cosmological implications and the quantization of the noncommutative action.

\bigskip
\subsection*{Acknowledgments}
\noindent 
The work of X.~C. was supported in part by a scholarship of the Universit\'e libre de Bruxelles. The work of A.~K. was partially supported by the GRDF grant 3305.

\bigskip


\end{document}